\documentclass[sigconf]{acmart}

\setcopyright{none}
\renewcommand\footnotetextcopyrightpermission[1]{}
\acmConference{}{}{}
\acmBooktitle{}
\acmYear{}
\copyrightyear{}
\acmDOI{}
\acmISBN{}

\AtBeginDocument{%
  }

\usepackage{tikz}
\usepackage{amsmath}
\usepackage{xurl}
\usepackage{hyperref}
\usepackage{multirow}  
\usepackage{xcolor}    
\usepackage{pifont}
\usetikzlibrary{patterns}
\usepackage{makecell}
\usepackage{tabularx}
\usepackage{listings}
\usepackage{booktabs}
\usepackage[most]{tcolorbox}
\newtcolorbox{findingbox}{
    enhanced,
    arc=1.5mm,                
    boxrule=0pt,              
    colback=gray!10!white,    
    frame hidden,             
    left=2mm,
    right=2mm,
    top=2mm,
    bottom=2mm,
    fontupper=\small,         
}

\definecolor{codegreen}{rgb}{0,0.6,0}
\definecolor{codegray}{rgb}{0.5,0.5,0.5}
\definecolor{codepurple}{rgb}{0.58,0,0.82}
\definecolor{backcolour}{rgb}{0.976, 0.976, 0.976}
\definecolor{keywordblue}{rgb}{0.13,0.13,1}
\definecolor{barred}{HTML}{FF7676}
\definecolor{baryellow}{HTML}{F9D662}
\definecolor{bargreen}{HTML}{7CAB7D}
\definecolor{barblue}{HTML}{75B7D1}

\lstdefinelanguage{json}{
    morekeywords={Name, Description, CodeSnippet, RiskLevel, StageTag}, 
    literate=
     *{0}{{{\color{codepurple}0}}}{1}
      {1}{{{\color{codepurple}1}}}{1}
      {2}{{{\color{codepurple}2}}}{1}
      {3}{{{\color{codepurple}3}}}{1}
      {4}{{{\color{codepurple}4}}}{1}
      {5}{{{\color{codepurple}5}}}{1}
      {6}{{{\color{codepurple}6}}}{1}
      {7}{{{\color{codepurple}7}}}{1}
      {8}{{{\color{codepurple}8}}}{1}
      {9}{{{\color{codepurple}9}}}{1}
      {:}{{{\color{black}{:}}}}{1}
      {,}{{{\color{black}{,}}}}{1}
      {\{}{{{\color{black}{\{}}}}{1}
      {\}}{{{\color{black}{\}}}}}{1}
      {[}{{{\color{black}{[}}}}{1}
      {]}{{{\color{black}{]}}}}{1},
}

\usepackage{xspace} 

\usepackage{pgfplots}
\pgfplotsset{compat=1.17} 

\newcommand{\systemname}{\textit{AutoBypass}\xspace} 
\newcommand{\boazname}{\textit{BOAZ}\xspace} 
\newcommand{\inceptorname}{\textit{Inceptor}\xspace} 
\newcommand{\dantename}{\textit{Dante-7B}\xspace} 

\begin{document}

\title{Mutate to Bypass: Autonomous Endpoint Evasion via Knowledge-Driven Multi-Agent Orchestration}

\author{Weifeng Yuan}
\authornote{Both authors contributed equally to this research.}
\orcid{0009-0000-4910-3041}
\affiliation{%
  \institution{Huazhong University of Science and Technology}
  \city{Wuhan}
  \country{China}
}
\email{cougarsecu@gmail.com}

\author{Wenbo Guo}
\authornotemark[1] 
\affiliation{%
\institution{Nanyang Technological University}
  \country{Singapore}
  }
\email{honywenair@gmail.com}
\orcid{0000-0001-6655-8179}

\author{Qingyun Du}
\affiliation{%
  \institution{Huazhong University of Science and Technology}
  \city{Wuhan}
  \country{China}
  }
\email{duqingyun@hust.edu.cn}
\orcid{0009-0001-7075-5113}

\author{Jun Chen}
\affiliation{%
  \institution{Huazhong University of Science and Technology}
  \city{Wuhan}
  \country{China}
}
\email{13507136963@139.com}
\orcid{0009-0002-4937-6466}

\author{Feng Dong}
\correspondingauthor
\affiliation{%
   \institution{Huazhong University of Science and Technology}
  \city{Wuhan}
  \country{China}
}
\email{dongfeng@hust.edu.cn}
\orcid{0000-0001-7091-2169}

\author{Haoyu Wang}
\affiliation{%
  \institution{Huazhong University of Science and Technology}
  \city{Wuhan}
  \country{China}}
\email{haoyuwang@hust.edu.cn}
\orcid{0000-0003-1100-8633}

\author{Yang Liu}
\affiliation{%
  \institution{Nanyang Technological University}
  \country{Singapore}}
\email{yangliu@ntu.edu.sg}
\orcid{0000-0001-7300-9215}

\begin{abstract}
While a wealth of evasion tactics is constantly released via threat intelligence, open-source repositories, and security blogs, a fundamental question remains: are state-of-the-art Endpoint Detection and Response (EDR) solutions actually resilient against publicly documented attack vectors? Rigorously answering this requires translating scattered security knowledge into operational payloads. However, the security field currently lacks a systematic approach to autonomously synthesize isolated tactics into functional executables, and the lack of transparency in EDR alerts prevents effective, automated payload refinement. Consequently, end-to-end automation of EDR resilience assessment remains a significant challenge.

To bridge this gap, we propose \systemname{}, a novel framework that formulates EDR bypass generation as an autonomous, knowledge-grounded, and closed-loop multi-agent workflow. At its core, a \textit{Detection-Aware Knowledge Base} (KB) processes raw threat intelligence, expert analyses, and open-source PoCs, converting them into a structured taxonomy of evasion methodologies and operational security constraints to anchor agentic reasoning. Leveraging the KB, a collaborative multi-agent architecture orchestrates high-level attack planning, polymorphic code generation, and binary compilation. Concurrently, a telemetry-driven alert reasoning engine diagnoses the underlying causes of execution failures, supplying actionable feedback to continuously evolve the evasion strategy. In a comprehensive evaluation against seven prominent commercial endpoint security platforms, \systemname{} successfully bypassed all targets, demonstrating peak evasion rates of 90\% against Windows Defender and 86.7\% against Trend Micro AV. Extensive ablation experiments highlight the critical role of the KB, which empowers smaller, open-weight language models (8B parameters) to improve their evasion success from a baseline of 27-53\% to 43-83\%, achieving performance parity with massive proprietary models.
\end{abstract}

\ccsdesc[500]{Security and privacy~Intrusion detection systems}
\ccsdesc[500]{Security and privacy~Malware and its mitigation}
\ccsdesc[300]{Security and privacy~Operating systems security}


\maketitle

\section{Introduction}
\label{sec:intro}

A single successful EDR evasion grants attackers persistent, invisible access to an entire enterprise network, enabling lateral movement, data exfiltration, and long-term control while remaining undetected by the security operations center~\cite{unit42_edr_bypass, vmray_edr_bypass}. Yet the security community continuously publishes techniques probing the limits of EDR defenses. Threat intelligence reports detail evasion techniques, proof of concept repositories demonstrate working exploits, and technical blogs dissect the internals of commercial defenses. This creates a fundamental tension, as the knowledge required for EDR evasion is openly available, but whether modern defenses can truly withstand these published techniques remains an open question.

Answering this question requires understanding the multi-layered nature of modern EDR defenses. Today's systems extend far beyond traditional signature matching, integrating kernel-level telemetry, memory scanning, and cloud-assisted behavioral profiling into a defense-in-depth architecture that scrutinizes every stage of the attack lifecycle~\cite{crowdstrike_ml, sentinelone_ml,Gartner_guide,alahmadi202299,osti10085663,milajerdi2019poirot,wang2020you}. For an adversary, this means that successful evasion now demands simultaneously evading static analysis of on-disk artifacts, behavioral monitoring of runtime execution, and heuristic inspection of memory patterns~\cite{hand2023evading, crowdstrike_dll_sideloading, bitdefender_dll_sideloading, kaspersky_dll_sideloading}. Evasion has thus transformed from a static packaging task into a dynamic, multi-dimensional orchestration problem. An attack must not only appear benign on disk, but also behave legitimately in memory and across the network.

Given this multi-dimensional challenge, can existing approaches provide a scalable solution? Manual techniques such as indirect system calls to bypass API hooks~\cite{junior2024hookchain} or repurposing EDR infrastructure for offensive operations~\cite{alachkar2025eviledr} have proven effective against specific targets, but they remain static and labor-intensive, requiring significant expertise to adapt to vendor updates. Automated approaches fare no better. Template-based frameworks like \boazname{}~\cite{boaz} and \inceptorname{}~\cite{inceptor} produce samples with recognizable fingerprints that EDRs quickly learn to detect, while domain-specific research like ANIMAGUS~\cite{zhou2023limits} targets only ransomware behaviors without addressing generic payload delivery \textbf{(Gap 1: Lack of Polymorphic Generation)}. Syntactic transformation methods, such as deterministic AST mutations~\cite{lepori2025automated}, can evade static signatures but cannot strategically pivot between offensive tradecrafts (e.g., switching from process injection to threadless execution) when behavioral monitoring blocks the original approach \textbf{(Gap 2: No Semantic-Level Adaptation)}. Furthermore, recent LLM-based tools function as passive coding assistants rather than autonomous agents~\cite{dante7b, anthropic_espionage, fang2024llm, deng2023masterkey}, and to our knowledge all existing methods operate in an open-loop manner: they lack a feedback mechanism to validate payloads against live defenses or interpret opaque EDR alerts, which typically flag a generic threat without disclosing the exact detection root cause. This lack of detail prevents the iterative refinement necessary to converge on successful evasions \textbf{(Gap 3: Absence of Closed-Loop Feedback)}.

We propose \systemname{}, which addresses these gaps through a knowledge-driven, feedback-directed reasoning process. Our approach centers on three key insights:


\noindent\textbf{Addressing Gap 1:} To overcome the lack of polymorphic generation, we introduce a multi-agent pipeline that actively randomizes and mixes alternative evasion implementations. It leverages a \textit{Detection-Aware Knowledge Base}, which distills threat intelligence and defensive rules~\cite{yara, sigma, elasticrules} into structured primitives. Acting as the active orchestrators, our agents intelligently select and compose these components. By dynamically substituting techniques and varying code structures under KB guidance, the multi-agent system breaks away from deterministic templates, achieving reliable polymorphism that raw LLMs cannot accomplish alone~\cite{anthropic_espionage}.

\noindent\textbf{Addressing Gap 2:} To achieve semantic-level adaptation, we employ a multi-agent architecture where a \textit{Strategist Agent} reasons over the KB to select and compose techniques based on detection semantics, not just syntactic patterns. This enables the system to pivot between fundamentally different offensive tradecrafts when one approach is blocked.

\noindent\textbf{Addressing Gap 3:} To close the feedback loop, a dedicated \textit{Tester Agent} executes samples in live EDR environments and infers detection root causes by correlating alert timing with observable system telemetry, including process activities, file operations, network traffic and Windows Event Logs~\cite{eventviewer}. These inferences feed back into the KB to drive iterative strategy evolution.

To realize these insights, \systemname{} integrates three core modules: a \textit{Detection-Aware Knowledge Base} that structures offensive knowledge, a \textit{Multi-Agent Sample Generation} pipeline that orchestrates strategy selection, code synthesis, and compilation, and an \textit{Alert Reasoning} module that interprets EDR feedback to drive iterative refinement. We emphasize that the key contribution of \systemname{} lies not in the discovery of fundamentally new evasion primitives, but rather in the \textbf{autonomous, knowledge-driven orchestration of known techniques} to achieve robust polymorphism and semantic adaptation.

We evaluate \systemname{} against seven industry-leading endpoint protection platforms. Our results demonstrate that the framework successfully evades all seven targets, achieving peak evasion rates of 90\% against Windows Defender and 86.7\% against Trend Micro AV. Ablation studies confirm the KB as the decisive factor, elevating computationally efficient, open-weight models (e.g., \texttt{Qwen3-8B}, \texttt{Llama-3.1-8B}) from a baseline of 27-53\% to 43-83\% evasion rates and closing the gap with large proprietary models. Furthermore, our analysis reveals that techniques leveraging trusted execution contexts, such as DLL sideloading, are particularly effective against defense-in-depth architectures, as they allow the payload to inherit the benign reputation of the host process.

In summary, we make the following contributions:
\begin{itemize}
\item We introduce \systemname{}, to our knowledge the first framework to model EDR evasion as a knowledge-driven, feedback-directed agentic process, automating the entire evasion lifecycle from technique selection to iterative refinement.
\item We design a Detection-Aware Knowledge Base that structures fragmented security knowledge into actionable representations, enabling small open-weight models (8B parameters) to match the performance of large proprietary models.
\item Through extensive evaluation against seven commercial EDRs, we demonstrate that current defenses remain vulnerable to polymorphic, knowledge-driven attacks, and identify trusted execution contexts as a critical blind spot in behavioral detection.
\end{itemize}

\section{Background}
\label{sec:background}

\subsection{EDR Detection Mechanisms}
\label{subsec:edr_mechanisms}

\begin{figure}[htbp]
    \centering
    \includegraphics[width=1\linewidth]{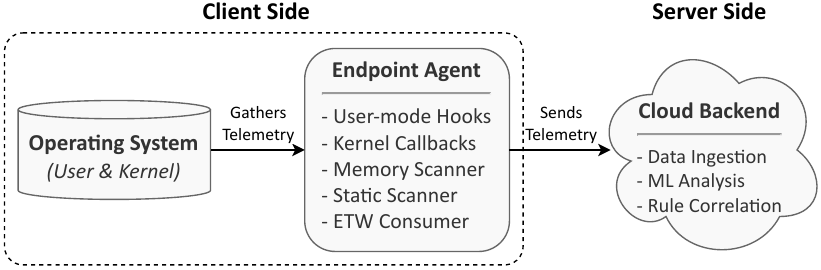}
    \caption{Overview of EDR Architecture}
    \label{fig:edr_archi}
\end{figure}

As shown in~\autoref{fig:edr_archi}, modern EDR solutions typically adopt a client-server architecture~\cite{hand2023evading}. The endpoint agent functions as a sensing array, capturing system telemetry, including process execution, file system operations, and network events. Crucially, these agents operate using a hybrid detection model: they employ local heuristics and static analysis to interdict immediate threats in real-time, while simultaneously transmitting telemetry to a central server for aggregate correlation and retrospective analysis. The agent itself utilizes a modular design composed of specialized sensors, each targeted at monitoring distinct operating system vectors.

These sensors collaborate to establish a multi-layered detection method. At the user level, EDRs inject hooks into critical libraries, such as \texttt{ntdll.dll}, to intercept API calls and identify suspicious behavioral patterns like process injection. To ensure visibility even if user-mode hooks are bypassed, EDRs utilize kernel-level callbacks (e.g., \texttt{PsSetCreateProcessNotifyRoutine}) and Event Tracing for Windows (ETW)~\cite{etw}. These mechanisms provide high-fidelity information regarding process, thread, and image loading events. Furthermore, to counter fileless threats that reside solely in RAM, EDRs employ memory scanners to periodically inspect the address space for anomalies, such as decrypted payloads or evidence of reflective DLL loading~\cite{Reflective_Code_Loading}.

This overlapping sensor coverage creates a robust defense-in-depth architecture~\cite{hassan2020tactical}. Consequently, effective evasion requires an integrated strategy rather than bypassing a single detection mechanism. The attacker must coordinate a sequence of actions where every step remains below the detection thresholds of static analysis, behavioral monitoring, and memory inspection simultaneously. Triggering even a single sensor can initiate correlation logic that reveals the entire attack chain~\cite{milajerdi2019holmes, han2020unicorn}.

Beyond runtime detection, EDRs leverage codified rules for pattern matching. Static signatures use YARA~\cite{yara} to identify malicious byte sequences, while behavioral rules in Sigma~\cite{sigma} and Elastic~\cite{elasticrules} format describe suspicious execution patterns. Complementing these deterministic rules, Operational Security (OPSEC) guidelines~\cite{opsec} encode negative constraints (e.g., ``avoid RWX memory'') that adversaries must satisfy to remain undetected~\cite{liao2016acing}.

\subsection{Taxonomy of Evasion Techniques}
\label{subsec:bypass_taxonomy}


To systematically understand EDR evasion, we focus on the shellcode loader because shellcode is universal in offensive operations. Standard C2 frameworks like Cobalt Strike and Sliver~\cite{cobaltstrike, sliver} generate payloads as position-independent shellcode, and mature translation tools such as Donut~\cite{donut} and sRDI~\cite{srdi} convert arbitrary PE files into shellcode format. By targeting the loader that transforms this raw shellcode into an executable state, we address a core delivery mechanism shared across diverse threats.

\begin{figure}[htbp]
    \centering
    \includegraphics[width=1\linewidth]{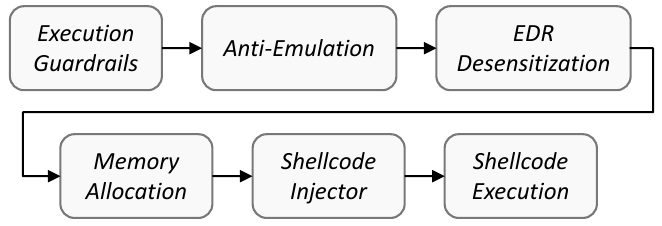}
    \caption{Overview of a Shellcode Loader}
    \label{fig:loader_lifecycle}
\end{figure}

As illustrated in~\autoref{fig:loader_lifecycle}, the execution lifecycle of a modern shellcode loader is modeled as six stages~\cite{dobin_edr}. This structured classification provides a theoretical foundation for systematic evasion automation. The process initiates with \textit{Execution Guardrails}, which validates environmental criteria to prevent premature execution. The loader then employs \textit{Anti-Emulation} tactics to delay execution or detect malware sandboxes. Prior to malicious activity, \textit{EDR Desensitization} attempts to blind sensors. Subsequently, \textit{Memory Allocation} reserves inconspicuous memory regions while avoiding suspicious permission combinations such as Read-Write-Execute (RWX) mappings that are commonly flagged by EDR systems. The \textit{Shellcode Injector} writes the decrypted payload into this allocated space. Finally, \textit{Shellcode Execution} transfers control to the payload. Across these stages, the techniques employed at each phase have evolved through a continuous adversarial cycle between attackers and defenders.

To facilitate understanding, \autoref{tab:bypass_evolution} summarizes the adversarial dynamics at each stage, illustrating how evasion techniques have evolved in response to specific detection challenges.


\begin{table}[htbp]
\centering
\caption{Evasion Techniques Across Loader Stages.}
\label{tab:bypass_evolution}
\footnotesize
\resizebox{\columnwidth}{!}{
\begin{tabular}{@{}lll@{}}
\toprule
\textbf{Stage} & \textbf{Detection Challenge} & \textbf{Evasion Technique} \\
\midrule
\textit{Execution Guardrails} & Sandbox analysis & Environment-based decryption \\
\textit{Anti-Emulation} & Accelerated sleep emulation & Computation-heavy delays \\
\textit{EDR Desensitization} & User-mode API hooks & ETW patching \\
\textit{Memory Allocation} & RWX memory flagging & File-backed DLL overwriting \\
\textit{Shellcode Injector} & Thread creation monitoring & Process hollowing \\
\textit{Shellcode Execution} & Thread anomaly detection & Windows fibers \\
\bottomrule
\end{tabular}
}
\end{table}

\section{Threat Model}
\label{sec:threat_model}

\noindent\textbf{Attacker's Goal.} Representing a critical post-exploitation phase where payload delivery dictates the success of an intrusion, the adversary aims to execute arbitrary shellcode on a target endpoint without triggering AV/EDR interdiction. Success is defined as the payload completing execution without being blocked or generating high-confidence alerts that would prompt immediate incident response.

\noindent\textbf{Attacker's Capabilities.} We assume the attacker has already achieved initial access (e.g., via phishing or other means) and holds standard user privileges on the compromised system. Within this environment, the adversary is capable of generating and compiling executable artifacts offline within their own infrastructure, and subsequently deploying the resulting payloads to the target endpoint.

\noindent\textbf{Attacker's Knowledge.} We adopt a gray-box model. The attacker lacks access to the proprietary source code, internal detection logic, or machine learning models of the target AV/EDR. However, the attacker can leverage publicly available resources, including open-source detection rules (e.g., YARA~\cite{yara}, Sigma~\cite{sigma}, Elastic rules~\cite{elasticrules}), technical blogs documenting evasion techniques, and OPSEC guidelines. Additionally, the attacker can observe AV/EDR feedback (alerts, blocked processes) and iteratively refine their approach based on this information.

\noindent\textbf{Scope.} We focus on the shellcode loader as the attack vector. The system takes arbitrary position independent shellcode as input and produces an evasive sample designed to execute this shellcode while evading AV/EDR detection. The output can be a standalone executable or a DLL designed for sideloading~\cite{yu2024file}. We do not consider exploitation of software vulnerabilities (e.g., memory corruption, zero-day exploits) or attacks targeting the AV/EDR agent itself (e.g., driver vulnerabilities, privilege escalation to disable AV/EDR).
\section{System Design}
\label{sec:system_design}

\begin{figure*}[htbp]
    \centering
    \includegraphics[width=1\linewidth]{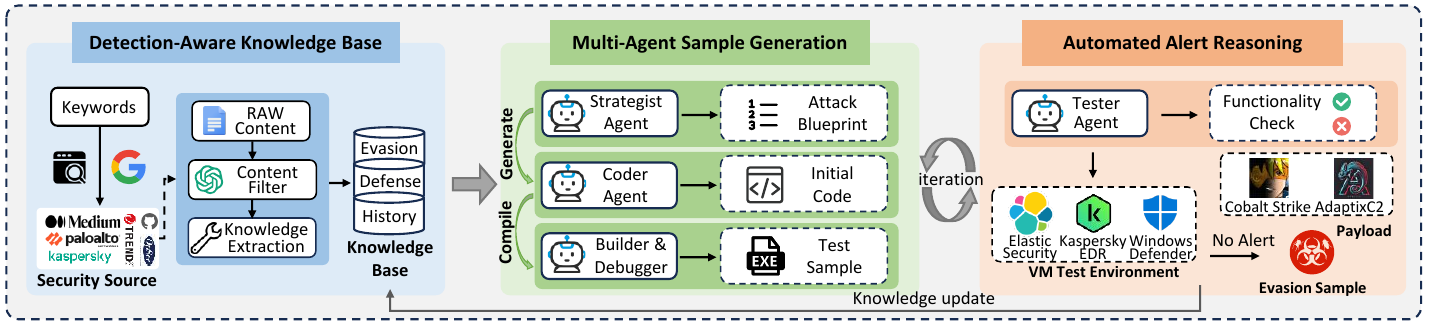}
    \caption{System Overview of \systemname}
    \label{fig:overview}
\end{figure*}

As illustrated in \autoref{fig:overview}, \systemname{} comprises three modules that form a closed-loop system: \textbf{(1)} The \textit{Detection-Aware Knowledge Base} structures evasion techniques, detection rules, and OPSEC guidelines into a unified representation for agentic reasoning. \textbf{(2)} The \textit{Knowledge-Driven Sample Generation} module employs a multi-agent architecture (\textit{Strategist}, \textit{Coder}, \textit{Builder}, \textit{Debugger}) to transform strategic plans into compilable executables. \textbf{(3)} The \textit{Automated Alert Reasoning} module executes samples against live EDRs, infers detection root causes, and feeds insights back to refine future iterations.

\subsection{Detection-Aware Knowledge Base}
\label{subsec:kb}

General-purpose LLMs are proficient at code generation but lack up-to-date, domain-specific security expertise, often generating unreliable or non-functional code for complex exploit chains~\cite{spracklen2025we}. This results in samples that are either functionally unstable or operationally insecure. However, successful autonomous evasion relies on coordinating three categories of critical information: (1) \textit{Evasion \& Encoding Techniques} that define concrete implementation methods for each stage, (2) \textit{Defensive Rules} that specify OPSEC constraints and behavioral patterns triggering detection, and (3) \textit{Historical Experience} that records the outcomes and failure causes of past runs. Currently, these sources are fragmented across heterogeneous repositories: evasion techniques are scattered in blogs and PoC repositories, detection rules reside in databases (YARA, Sigma, Elastic), and OPSEC guidelines exist as unstructured advisory documents. Prior work has demonstrated the value of extracting structured knowledge from such unstructured reports for malware detection~\cite{guo2026bridging, liao2016acing}. We extend this paradigm to the offensive domain.

To bridge this gap, we construct a \textit{Detection-Aware Knowledge Base (KB)} that organizes these three categories into structured, queryable functional modules. We first describe the KB's construction from heterogeneous sources (\S\ref{subsubsec:kb_construction}), then detail how each category is formally represented (\S\ref{subsubsec:knowledge_dimensions}).

\subsubsection{Knowledge Base Construction}
\label{subsubsec:kb_construction}

The core challenge of KB construction lies in transforming security knowledge scattered across heterogeneous sources into a structured, queryable representation. We address this through a three-stage processing pipeline: \ding{172} data source identification, \ding{173} data collection, and \ding{174} knowledge extraction.

\noindent \textbf{\ding{172} Data Source Identification.}
We target two categories of external data sources. For offensive primitives, the relevant sources include technical blogs from security researchers, threat analysis reports from security vendors, and proof of concept repositories on GitHub~\cite{timeline, Encryptions}. For defensive constraints, we target detection rule databases (YARA~\cite{yara}, Sigma~\cite{sigma}, Elastic~\cite{elasticrules}) and OPSEC advisory documents~\cite{opsec}. To locate these sources, we employ the Google Search API with domain-specific terms~\cite{guo2024packageintel} such as ``shellcode loader,'' ``process injection,'' and ``EDR bypass.''

\noindent \textbf{\ding{173} Data Collection.}
For web pages identified through search, a custom crawler retrieves the HTML content and extracts the main text body, filtering out navigation elements and advertisements. For GitHub repositories, we use the GitHub API to clone the entire project and parse both README documentation and source code files. For detection rule repositories, we directly clone the official repositories, which maintain well-defined directory structures.


\noindent \textbf{\ding{174} Knowledge Extraction.}
The collected raw data requires different extraction strategies based on its structure. For unstructured sources (blogs, reports, OPSEC documents), we employ a two-stage extraction pipeline. First, a lightweight model performs relevance filtering to discard off-topic content and retain only security-relevant passages. Second, an LLM-based extraction agent parses the filtered content into structured knowledge. As shown in~\autoref{fig:extraction_prompts}, the agent uses task-specific prompts to extract offensive techniques into formalized JSON tuples and summarize OPSEC guidelines into concise negative constraints. All extracted code snippets undergo manual expert review to ensure compilability and functional correctness. This expert oversight, combined with our strict restriction to highly credible sources, ensures that untrusted data does not compromise the validity of the final outputs. For structured detection rules, we develop Python-based parsers that directly ingest YARA, Sigma, and Elastic rules. To filter redundancy, we leverage directory structures to retain only Windows-specific malware rules and apply keyword filtering on Sigma rules to select those relevant to offensive tradecrafts (e.g., process injection, AMSI bypassing).

In our implementation, we collected 1351 web pages through Google Search API using 20 domain-specific keywords, along with 6 GitHub repositories. We employed \texttt{GPT-4o} to filter out off-topic content, reducing the corpus to 427 relevant passages. The LLM-based extraction agent, powered by \texttt{GPT-5.1}, parsed these passages into structured entries. The final KB contains 19 AV/EDR evasion techniques, 5 encoding schemes, 68 OPSEC constraints, and 1537 detection rules (407 YARA, 400 Sigma, 730 Elastic). The hybrid storage is implemented using MySQL for the relational database and Chroma for the vector database, with embeddings generated by the \texttt{all-MiniLM-L6-v2} model. All extracted code snippets were reviewed by 4 security experts.


\begin{figure}[htbp]
\centering
\begin{tcolorbox}[colback=gray!5!white, colframe=black,
coltitle=white, colbacktitle=black,
sharp corners=south, boxrule=0.6pt, arc=2mm,
width=1\linewidth ,title=\textbf{Prompts for Knowledge Extraction}]
\small
\textbf{Task 1: Offensive Technique Extraction}

\textit{Extract the C++ code snippet for the described injection technique and classify its execution stage. Format the output strictly as a JSON tuple: T=⟨Name, Description, CodeSnippet, RiskLevel, StageTag⟩} \vspace{0.1cm} 

\textbf{Task 2: OPSEC Constraint Summarization} 

\textit{Summarize the following OPSEC advice into a concise constraint (e.g., `Do not use API X in context Y').}
\end{tcolorbox}
\caption{Prompts used for knowledge construction}
\label{fig:extraction_prompts}
\end{figure}

\subsubsection{Knowledge Organization}
\label{subsubsec:knowledge_dimensions}

To serve diverse query patterns during generation, the KB adopts a hybrid storage architecture. The \textit{Strategist} requires both structured filtering (e.g., selecting techniques by execution stage and risk level) and semantic retrieval (e.g., finding techniques similar to a given description). The \textit{Coder} retrieves OPSEC constraints to guide stealthy synthesis and historical records enable learning from past failures. A relational database (RDB) stores structured metadata for precise queries, while a vector database (VDB) stores semantic embeddings for fuzzy retrieval. As detailed in~\autoref{tab:kb_schema}, this design enables compound queries such as: \textit{``Retrieve techniques where Stage is Payload Execution (RDB filter) AND Description is semantically similar to Threadless (VDB similarity).''}

\begin{table}[htbp]
\centering
\caption{Schema for the Hybrid Knowledge Base}
\label{tab:kb_schema}
\resizebox{\columnwidth}{!}{%
\begin{tabular}{@{}lll@{}}
\toprule
\textbf{Data} & \textbf{Storage\textsuperscript{*}} & \textbf{Schema} \\ \midrule
\multirow{2}{*}{\textbf{Evasion \& Encode}} & RDB & \texttt{id, name, snippet, risk, stage} \\ \cmidrule(l){2-3}
 & VDB & \texttt{embedding} $\leftarrow$ \texttt{desc}; \texttt{meta: \{id\}} \\ \midrule
\multirow{2}{*}{\textbf{Defensive Rules (OPSEC)}} & RDB & \texttt{rule\_id, type, content} \\ \cmidrule(l){2-3}
 & VDB & \texttt{embedding} $\leftarrow$ \texttt{constraint}; \texttt{data: constraint} \\ \midrule
\textbf{History} & RDB & \texttt{run\_id, blueprint, result, reason} \\ \bottomrule
\end{tabular}%
}
\vspace{4mm}
\parbox{\linewidth}{\footnotesize \textsuperscript{*}RDB: Relational Database, VDB: Vector Database.}
\end{table}

\noindent \textbf{Evasion \& Encoding Techniques.}
This component serves as the structured arsenal for sample generation. Each technique $T$ is modeled as a formalized tuple:
\begin{center}
\resizebox{\columnwidth}{!}{%
$T = \langle \mathit{Name}, \mathit{Description}, \mathit{CodeSnippet}, \mathit{RiskLevel}, \mathit{StageTag} \rangle$%
}
\end{center}

The \textit{RiskLevel} (Low/Medium/High) quantifies detection sensitivity, while the \textit{StageTag} anchors the technique to a specific phase of the shellcode loader lifecycle (detailed in~\S\ref{subsec:bypass_taxonomy}). \autoref{lst:fiber_example} illustrates a concrete entry. Additionally, this category includes encoding schemes (e.g., XOR, AES) paired with decryption stubs to obfuscate static signatures. To ensure generated blueprints are valid and stealthy, we enforce three classes of logic constraints: \textit{(1) Intra-stage Exclusivity} prevents redundant technique combinations within the same stage; \textit{(2) Inter-stage Synergy} promotes combinations that collectively reduce detection likelihood; and \textit{(3) Environmental Dependencies} models preparatory techniques (e.g., \texttt{DLL Sideloading}) as modifiers that can co-exist with other techniques.

\begin{lstlisting}[language=json, caption={Fiber-based shellcode execution entry (truncated).}, label={lst:fiber_example}]
{
  "Name": "Fiber Execution",
  "Description": "Executes the payload by converting the current thread to a fiber and scheduling a new fiber that transfers control to the shellcode.",
  "CodeSnippet": "void WINAPI FiberFunction(PVOID lpParameter) { ... } ... SwitchToFiber(payloadFiber);", 
  "RiskLevel": "Medium",
  "StageTag": "Shellcode Execution"
}
\end{lstlisting}

\noindent \textbf{Defensive Rules.}
This category equips agents with defensive awareness and comprises two components. \textit{OPSEC Constraints} are injected as system prompts to guide the \textit{Coder} in generating stealthy implementations. \textit{Detection Rules} (YARA/Sigma/Elastic) are utilized by the \textit{Builder} to proactively scan generated samples, filtering out binaries that trigger known static signatures or behavioral heuristics before deployment.

\noindent \textbf{Historical Experience.}
This module serves as the system's long-term memory, archiving iteration data as $\langle \textit{Blueprint}, \textit{Result}, \textit{Reason} \rangle$ triplets. The \textit{Reason} field captures the inferred cause of failure (e.g., ``Memory Scanning'') derived from alert telemetry. This data drives evolutionary refinement, allowing the \textit{Strategist} to prune ineffective strategies and pivot to counter-measures in subsequent iterations.

\subsection{Multi-Agent Sample Generation}
\label{subsec:sample_generation}

This module operationalizes the KB through a multi-agent architecture built upon the LangGraph framework~\cite{langgraph}. We decompose the sample generation task into four specialized roles: the \textit{Strategist} formulates high-level attack plans by querying KB knowledge, the \textit{Coder} synthesizes polymorphic C++ source code guided by OPSEC constraints, the \textit{Builder} compiles the source and performs defensive pre-checks, and the \textit{Debugger} automatically recovers from compilation failures. These agents operate in a sequential pipeline, with the workflow state passed from \textit{Strategist} to \textit{Coder} to \textit{Builder}, and conditionally looping back through \textit{Debugger} when errors occur.

\subsubsection{Strategist Agent}
\label{subsubsec:strategist_agent}

The \textit{Strategist} is responsible for synthesizing high-level attack plans that define the architecture of a shellcode loader. Its output is a structured JSON object termed \textit{Blueprint}, which specifies the evasion technique for each execution stage, the payload encoding scheme, compiler flags, and output format. \autoref{tab:blueprint_structure} details the blueprint schema.

\begin{table}[htbp]
\centering
\caption{Structure of the \textit{Blueprint} JSON}
\label{tab:blueprint_structure}
\small
\begin{tabularx}{\linewidth}{@{} l X @{}}
\hline
\textbf{Key} &  \textbf{Description} \\ \hline
\textbf{Rationale}  & Explains the strategic reasoning \\
\textbf{Evasion\_tech}  & Specifies the evasion techniques \\
\textbf{Encode\_tech}  & Defines the payload obfuscation scheme \\
\textbf{Compiler\_flags} & Sets compiler optimizations \\
\textbf{Sample\_type} & Determines the sample format (e.g., exe or dll) \\ \hline
\end{tabularx}
\end{table}

To generate each blueprint, the \textit{Strategist} queries two knowledge categories from the KB. First, it retrieves \textit{Evasion \& Encoding Techniques} from the relational database, filtering candidates by \textit{StageTag} to ensure coverage of all execution stages and by \textit{RiskLevel} to balance stealth against implementation complexity. The agent strictly enforces the logic constraints defined in~\S\ref{subsubsec:knowledge_dimensions}: intra-stage exclusivity ensures only one technique per stage, inter-stage synergy promotes complementary combinations, and environmental dependencies are correctly modeled.

Second, the \textit{Strategist} leverages \textit{Historical Experience} to guide iterative refinement. When previous attempts have failed, the agent retrieves the recorded failure causes (e.g., ``Memory Scanning'') and performs a semantic search over the vector database to identify relevant countermeasures. For instance, a memory scanning failure prompts retrieval of techniques such as \textit{Sleep Obfuscation} or \textit{Stack Spoofing}. This feedback-driven mechanism enables the system to systematically explore the technique space while avoiding previously ineffective combinations. Crucially, rather than executing predefined database lookups, the \textit{Strategist} employs semantic reasoning to translate opaque failure phenomena into conceptual counter strategies, demonstrating a dynamic decision process that rigid programmatic algorithms cannot replicate.

\subsubsection{Coder Agent}
\label{subsubsec:coder_agent}

The \textit{Coder} receives the blueprint from the \textit{Strategist} and synthesizes fully functional C++ source code. We employ LLM-driven code generation rather than template-based approaches, which naturally introduces polymorphism through variations in variable naming, control flow structure, and code organization, thereby resisting hash-based static detection.


To ensure both functional correctness and operational stealth, the \textit{Coder} queries two knowledge categories from the KB. First, it retrieves the \textit{CodeSnippet} field from \textit{Evasion \& Encoding Techniques} via exact lookup in the relational database, using the technique names specified in the blueprint. These verified snippets serve as implementation references for complex operations such as indirect syscalls or memory manipulation, reducing the risk of functional errors. Second, the \textit{Coder} performs a semantic search over \textit{OPSEC Constraints} in the vector database, using the selected technique names as query keys to retrieve relevant guidelines (e.g., ``Avoid RWX memory allocations''). These constraints are injected into the generation prompt to guide the LLM toward stealthy implementations.

\subsubsection{Builder and Debugger Agents}
\label{subsubsec:builder_agent}

The \textit{Builder} receives the C++ source code from the \textit{Coder} and produces a compilable executable (EXE or DLL). It invokes the MSVC toolchain with the compiler flags specified in the blueprint. If compilation fails, the \textit{Debugger} analyzes the error messages and automatically modifies the source code to resolve issues such as missing headers, type mismatches, or syntax errors. This self-repair loop continues until compilation succeeds or a maximum retry limit is reached.

After successful compilation, the \textit{Builder} performs a defensive pre-check using \textit{Defensive Rules} from the KB. It scans the compiled binary against YARA signatures to detect known malicious byte patterns, and analyzes the source code's API call sequences against Sigma and Elastic rules to identify behavioral patterns that may trigger runtime detection. Samples matching known signatures are discarded, ensuring only proactively sanitized payloads proceed to the evaluation phase.

\subsubsection{Implementation Details}
\label{subsubsec:generation_impl}

\begin{figure*}[htbp]
\centering
\begin{tcolorbox}[colback=gray!5!white, colframe=black,
  coltitle=white, colbacktitle=black,
  sharp corners=south, boxrule=0.8pt, arc=2mm,
  width=1\linewidth ,title=\textbf{Prompts for Generation Agents}]
\small

\textbf{Strategist Agent} \\
\begingroup \itshape 
``You are an elite Red Team Strategist specializing in EDR evasion. Your task is to formulate the optimal Attack Blueprint for the next iteration based on historical feedback.

\textbf{Capabilities:}

1. \textbf{Failure Analysis:} Diagnose the root cause of previous failures using EDR alerts or error logs (e.g., Static Detection vs. Memory Scanning).

2. \textbf{Knowledge Retrieval:} Query the Knowledge Base for compatible `evasion\_tech' and `encode\_tech' primitives.

3. \textbf{Evolutionary Logic:} Learn from `Evaluation History' to avoid known bad behavior.

\textbf{Decision Logic:}

- \textbf{First Run:} Prioritize low-risk, high-compatibility primitives.

- \textbf{Subsequent Iteration:} If the previous attempt failed, pivot to a counter-strategy (e.g., enable Sleep Obfuscation if Memory Scanning was detected).

\textbf{Output Constraint:} Return ONLY the structured `Blueprint' JSON object (rationale, evasion\_tech, encode\_tech, compiler\_flags, sample\_type). No markdown, no commentary.''
\endgroup 

\vspace{0.3cm}

\textbf{Coder Agent} \\
\begingroup \itshape 
``You are an expert C++ Malware Developer. Your task is to synthesize a high-variance shellcode loader based on the provided Blueprint.

\textbf{Workflow:}

1. \textbf{Retrieve:} Fetch verified code snippets for the selected techniques from the Knowledge Base.

2. \textbf{Constrain:} Strictly adhere to the injected OPSEC Constraints:
   $\langle OPSEC\_INSERT \rangle$
   (e.g., ``Do not use RWX memory,'' ``Use Indirect Syscalls for allocation'')
   
3. \textbf{Polymorphize:} Generate code with high structural variability (e.g., randomized variable names, control flow flattening) to evade static signatures.

\textbf{Hard Requirements:}
- Output ONLY raw, compilable C++ code (MSVC compatible).
- Do NOT use placeholders; ensure all logic is fully implemented.
- Do NOT include markdown fences or comments.''
\endgroup 

\end{tcolorbox}
\caption{Prompts for the \textit{Strategist} and \textit{Coder} Agents}
\label{fig:agent_prompts}
\end{figure*}

The multi-agent pipeline is implemented using the LangGraph framework~\cite{langgraph}, which models the workflow as a directed graph with agents as nodes and state transitions as edges. The workflow state is a structured object containing the current blueprint, generated source code, compilation status, and error messages. The pipeline executes sequentially through three stages: the \textit{Strategist} first generates a blueprint, then the \textit{Coder} produces source code, and finally the \textit{Builder} compiles and validates the output. If compilation fails, the \textit{Builder} attempts self-repair for up to 5 iterations. If all repair attempts fail, the workflow returns to the \textit{Coder} with the error context for code regeneration. After 5 consecutive \textit{Coder} failures, the system falls back to the \textit{Strategist} for re-planning, assuming the failure is strategic rather than syntactic. Each agent performs a functionally distinct and non-substitutable role. Removing any single agent breaks the pipeline entirely. \autoref{fig:agent_prompts} presents the prompts used for the \textit{Strategist} and \textit{Coder} agents.

All agents are powered by \texttt{GPT-5.1} operating at the default temperature to facilitate natural code variance and polymorphism. To strictly ensure output accuracy within this setup, the \textit{Strategist} outputs are constrained to structured JSON, while the \textit{Coder} is instructed to produce raw, compilable C++ code without markdown formatting.

The \textit{Builder} integrates with Microsoft Visual Studio Build Tools 2022. It executes \texttt{vcvars64.bat} to configure the MSVC environment and invokes \texttt{cl.exe} for compilation with the flags specified in the blueprint. Each compilation runs in an isolated environment to prevent interference between iterations. All inter-agent communication uses structured outputs (JSON blueprints, C++ source), with comprehensive logging of every iteration to support post-hoc analysis and reproducibility.


\subsection{Automated Alert Reasoning}
\label{subsec:closed_loop}

Generated samples are not guaranteed to evade detection on the first attempt, necessitating an iterative refinement process. However, when an EDR blocks a sample, the specific cause is often opaque: alerts may indicate only that a threat was detected, without specifying whether the trigger was a static signature, behavioral heuristic, or memory scan. As illustrated in~\autoref{tab:alert_granularity}, the level of diagnostic detail varies significantly across EDR products. To enable automated iteration, this module performs two tasks: a \textit{Tester} validates samples in a controlled environment and infers detection root causes from observable system telemetry, while an \textit{Evolutionary Update} mechanism commits these inferences to the KB's \textit{Historical Experience}, enabling the \textit{Strategist} to refine future blueprints.

\begin{table}[htbp]
  \centering
  \caption{Variability in EDR Alert Specificity}
  \label{tab:alert_granularity}
  \small 
  
  \resizebox{\linewidth}{!}{\begin{tabular}{@{} l c c c c @{}} 
    \toprule
    \textbf{AV/EDR} 
    & \textbf{\makecell[c]{Process \&\\ File Related}} 
    & \textbf{\makecell[c]{Engine\\ Attribution}} 
    & \textbf{\makecell[c]{Rule\\ Detail}} 
    & \textbf{\makecell[c]{Alert\\ Detail Level}} \\
    \midrule
    Windows Defender    & \checkmark & $\times$ & $\times$ & brief \\
    Trend Micro AV      & \checkmark & $\times$ & $\times$ & brief \\
    Avira               & \checkmark & $\times$ & $\times$ & brief \\
    McAfee              & \checkmark & $\times$ & $\times$ & brief \\
    Kaspersky EDR       & \checkmark & \textasciitilde & $\times$ & normal \\
    Bitdefender EDR     & \checkmark & \textasciitilde & $\times$ & normal \\
    Elastic Security    & \checkmark & \checkmark & \textasciitilde & verbose \\
    \bottomrule
  \end{tabular}
  }
  
  \vspace{1mm}
  \parbox{\linewidth}{\scriptsize
    \textit{Legend}: \checkmark (Provided), \textasciitilde (Partial), $\times$ (Not Provided). Process \& File Related: Identifies the malicious process and file. Engine Attribution: Specifies the detecting component (e.g., static, behavioral, memory, sandbox analysis). Rule Detail: Provides the specific rule content.
  }
\end{table}

\subsubsection{Tester Agent}
\label{subsubsec:tester_agent}

The \textit{Tester} receives the compiled executable from the \textit{Builder} and produces a structured test result containing the execution outcome and, if detection occurs, the inferred root cause. This agent operates in a two-phase validation process. First, in a baseline environment with the EDR disabled, it verifies that the sample functions correctly (e.g., establishes C2 connection, executes commands). Only samples passing this sanity check proceed to the second phase, where they are deployed against the active EDR to evaluate evasion capability.

When a sample is blocked, EDR alerts are often opaque. To enable actionable feedback, the \textit{Tester} applies heuristic inference by correlating observable system telemetry, including file existence, process lifetime, and network connectivity, to deduce the detection mechanism. As detailed in~\autoref{tab:alert_inference}, distinct behavioral patterns map to specific root causes: immediate file deletion indicates static signature matching, rapid process termination suggests behavioral blocking, failed C2 connection despite process survival points to network interception, and delayed termination after extended execution implies memory scanning.

\begin{table}[htbp]
\centering
\caption{Inference for EDR Alerts}
\label{tab:alert_inference}
\small
\renewcommand{\arraystretch}{0.6} 
\renewcommand{\tabularxcolumn}[1]{m{#1}} 
\begin{tabularx}{\linewidth}{@{} >{\raggedright\arraybackslash}X >{\raggedright\arraybackslash}X @{}}
\toprule
\textbf{Phenomena} & \textbf{Alert Reasoning} \\ \midrule
Sample is deleted or access denied upon disk write & \textbf{Static Detection:} Sample matches known signatures\\ \midrule
Process starts successfully but terminates within seconds & \textbf{Behavioral Blocking:} Runtime heuristics flagged the suspicious execution chain \\ \midrule
Process remains active but fails to establish C2 connection & \textbf{Network Interception:} Outbound traffic blocked by firewall or reputation filters \\ \midrule
Process runs for minutes before facing delayed termination & \textbf{Memory Scanning:} Periodic scan detected unbacked executable memory \\ \bottomrule
\end{tabularx}
\end{table}

For instance, if a sample survives initial execution but terminates after several minutes, the \textit{Tester} infers that memory scanning detected the decrypted payload. This structured inference transforms opaque EDR alerts into actionable feedback that drives the evolutionary update process.

\subsubsection{Evolutionary Update}
\label{subsubsec:evolutionary_update}

After each test iteration, the \textit{Tester} commits a structured record to the KB's \textit{Historical Experience}. Each record contains the target EDR, the complete blueprint, the test outcome, and for failed attempts, the inferred root cause. This persistent storage enables the system to accumulate operational experience across iterations.

The \textit{Strategist} queries this historical data to close the feedback loop. When generating a new blueprint, it first retrieves records matching the current target EDR. Successful blueprints serve as reference templates, biasing the agent toward proven technique combinations. Failed blueprints, paired with their inferred causes, act as negative constraints: the \textit{Strategist} avoids repeating the same technique selections that previously triggered detection. Furthermore, specific failure causes trigger targeted countermeasures. For instance, a \textit{Memory Scanning} failure prompts the \textit{Strategist} to incorporate defensive techniques such as sleep obfuscation or stack spoofing, while a \textit{Behavioral Blocking} failure leads to pivoting from monitored APIs (e.g., \texttt{CreateRemoteThread}) to less scrutinized alternatives (e.g., \texttt{NtCreateThreadEx}). This feedback-driven mechanism enables the system to systematically explore the evasion space while converging toward effective strategies.

\subsubsection{Implementation Details}
\label{subsubsec:implementation_details}

We implement the testing infrastructure using VMware Workstation to manage isolated VMs, with a Flask-based web service on the host and a Python agent inside each guest. The agent downloads and executes samples, then monitors system state at 2-second intervals. A successful evasion is defined as the payload establishing a stable C2 connection and maintaining heartbeat communication for over 30 seconds. For failed attempts, the agent infers root causes based on observable phenomena: file deletion within 5 seconds indicates static detection, process termination within 30 seconds suggests behavioral blocking, process survival without C2 connection implies network interception, and termination after 2 minutes points to memory scanning. Historical Experience records are stored in the same MySQL database as the KB, with each record containing target EDR, blueprint, outcome, and inferred cause. The system reverts VMs to clean snapshots after each test to eliminate cached signatures.
\section{Evaluation}
\label{sec:evaluation}

To demonstrate the efficacy and robustness of our autonomous EDR Evasion framework, we conducted a comprehensive evaluation. Our experiments were designed to answer the following Research Questions (RQs):

\begin{itemize}
    \item \textbf{RQ1 (Effectiveness):} How effective is \systemname{} at evading commercial EDR solutions, and how does it compare to existing approaches?
    \item \textbf{RQ2 (Ablation Study):} What is the contribution of each KB component to the evasion success?
    \item \textbf{RQ3 (Model Sensitivity):} How does performance vary across LLMs of different scales, and can smaller models achieve comparable results with KB support?
    \item \textbf{RQ4 (Analysis of Evasion Techniques):} Which evasion techniques and execution contexts are most effective against modern EDR defenses?
\end{itemize}

\begin{table*}[htbp]
\centering
\caption{Effectiveness of \systemname{} (with \texttt{GPT-5.1})}
\label{tab:rq1_effectiveness}
\resizebox{\textwidth}{!}{
    \begin{tabular}{@{} l l c c c c c c c @{}}
    \hline
    \textbf{Payload} & \textbf{Metric\textsuperscript{*}} & \textbf{W} & \textbf{T} & \textbf{K} & \textbf{A} & \textbf{M} & \textbf{B} & \textbf{E} \\ \hline
    \multirow{3}{*}{Cobalt Strike}
        & Success Evasion       & 90.0\%       & 86.7\%     & 53.3\%     & 23.3\% & 83.3\% & 13.3\%      & 16.7\%      \\
        & Avg. Time (min)    & 1.29       & 1.71       & 2.96       & 5.85 & 1.99 & 9.22        & 9.40        \\
        & Avg. Token (In/Out) & 7.2k/2.3k  & 9.5k/3.1k  & 15.2k/4.5k & 31.9k/11.4k & 8.6k/2.9k & 70.1k/18.1k & 51.9k/12.7k \\ \hline
    \multirow{3}{*}{MVP}
        & Success Evasion       & 86.7\% & 86.7\% & 73.3\% & 60.0\% & 83.3\% & 60.0\%   & 23.3\%  \\
        & Avg. Time (min)    & 1.25 & 1.58 & 1.54 & 2.40 & 1.49 & 2.00      & 4.96 \\
        & Avg. Token (In/Out) & 8.1k/2.3k & 9.6k/4.2k & 9.7k/3.1k & 13.1k/5.6k & 9.8k/4.1k & 13.3k/5.3k & 32.0k/11.6k \\ \hline
    \multirow{3}{*}{AdaptixC2}
        & Success Evasion       & 86.7\% & 83.3\% & 40.0\% & 46.7\% & 86.7\% & 76.7\% & 20.0\%  \\
        & Avg. Time (min)    & 1.28 & 1.68 & 3.02 & 2.86 & 1.44 & 1.52 & 6.66  \\
        & Avg. Token (In/Out) & 8.2k/2.2k & 10.1k/4.1k & 18.7k/6.6k & 15.5k/5.5k & 9.2k/3.6k & 9.4k/2.8k & 37.1k/14.7k \\ \hline
    \end{tabular}
}
\vspace{2mm}
\parbox{\linewidth}{\scriptsize
    \textsuperscript{*}W: Windows Defender, T: Trend Micro, K: Kaspersky, A: Avira, M: McAfee, B: Bitdefender, E: Elastic. Evasion rates are evaluated across 30 runs per AV/EDR and payload configuration ($N = 30$ per cell). Avg. Time and Avg. Token represent the average time and token cost \textbf{per successful evasion}.
}
\end{table*}

\subsection{Experimental Setup}
\label{subsec:setup}

\noindent \textbf{Testbed.}
All experiments were conducted in an isolated environment. The host machine runs Windows 11 25H2 and orchestrates virtual machines through VMware Workstation. The functional check VM runs Windows 10 22H2 with EDR disabled. To demonstrate the generalizability of our framework across different environments, the evaluation VMs were diversified across three distinct Windows OS versions. Specifically, Windows Defender AV (Engine: 1.1.25110.1)~\cite{windef} and TrendMicro PC-cillin AV (17.9.1106)~\cite{trendmicro} were evaluated on Windows 10 22H2. Avira Internet Security (1.1.115.3317)~\cite{avira}, McAfee LiveSafe (1.39.160.1)~\cite{mcafee}, and Kaspersky EDR (12.11.0.637)~\cite{kaspersky} were deployed on Windows 11 23H2. Finally, Bitdefender GravityZone EDR (7.9.29.589)~\cite{bitdefender} and Elastic Security (9.2.4)~\cite{elastic} were tested on Windows 11 24H2. All security products were evaluated using default installation settings.


\noindent \textbf{Evaluation Protocol.}
For our primary evaluation (RQ1), we utilized \texttt{GPT-5.1} across three distinct payloads: a standard Cobalt Strike shellcode of 926 bytes to assess evasion of established signatures~\cite{talos_cobaltstrike_2020}, a custom Minimum Viable Product (MVP) beacon jointly developed with \texttt{Claude Code (Claude Sonnet 4.5)} to isolate loader effectiveness from known payload patterns, and the AdaptixC2 framework~\cite{unit42_adaptixc2_2025} to confirm generalizability across modern threats. For each target, we conducted \textbf{30} iterations, where an iteration terminates upon the first successful evasion, defined as establishing a stable C2 connection without triggering EDR alerts, or after a maximum of five refinement attempts.

For the ablation (RQ2), sensitivity (RQ3), and technique analysis (RQ4), we selected representative subsets of targets to manage computational overhead. Specifically, RQ2 focuses on two distinct tiers of defense, Windows Defender and Kaspersky, while RQ3 and RQ4 evaluate a five target subset consisting of Defender, Trend Micro, Kaspersky, Bitdefender, and Elastic.

\subsection{Baseline Selection}
\label{subsec:baselines}

We compare \systemname{} against three representative approaches: (1) \boazname{}~\cite{boaz}, a template-based framework that combines predefined evasion modules; (2) \inceptorname{}~\cite{inceptor}, another template-based tool that generates shellcode loaders with configurable obfuscation; and (3) \dantename{}~\cite{dante7b}, a 7B-parameter LLM fine-tuned on malware development data. These baselines represent the current state-of-the-art in both template-based and LLM-based sample generation. We excluded Leucism~\cite{lepori2025automated} as its source code is unavailable, and its AST-based approach operates at the syntactic level without the semantic reasoning required to pivot between offensive tradecrafts.

\subsection{RQ1: Effectiveness}
\label{subsec:rq1_effectiveness}


\autoref{tab:rq1_effectiveness} summarizes the evasion success rates, average iteration time, and token cost across all seven targets and three payloads. The results demonstrate that \systemname{} consistently generates evasion samples against a majority of the tested endpoints. Specifically, the framework achieved consistently high success rates against Windows Defender (86.7\%--90.0\%) and Trend Micro (83.3\%--86.7\%) across all payload types. Notably, while heavily profiled payloads like Cobalt Strike struggled against Bitdefender (13.3\%), our framework effectively adapted when deploying MVP (60.0\%) and AdaptixC2 (76.7\%), demonstrating the system's robust capability to generalize across different malware families.


However, the system faced consistent challenges across all payloads with Elastic Security (16.7\%--23.3\%), which employs highly aggressive payload execution monitoring. Although Elastic open-sources its behavioral detection rules and YARA signatures, its static malware prevention engine~\cite{elastic_defend_policy, elastic_malicious_file} remains proprietary and unpublished. For Bitdefender, the specific difficulty observed with Cobalt Strike stems from its rigorous pre-execution sandbox, which terminates highly recognizable malicious process creation attempts before our evasion logic can initialize, a constraint largely mitigated when using alternative payloads.



To contextualize our results against existing literature, we compared \systemname{} against three baselines. For a fair comparison, we evaluated the widely adopted Cobalt Strike payload across static template frameworks (\boazname{} and \inceptorname{}) and a fine-tuned model (\dantename{}). As shown in \autoref{tab:sota_comparison}, \systemname{} significantly outperforms all baselines. Template-based tools suffer from deterministic code patterns that are easily fingerprinted, while \dantename{}, despite being fine-tuned on malware development data, achieved only a 4.7\% compilation success rate (14/300 attempts), highlighting the difficulty of generating functional exploit code without structured knowledge guidance. In contrast, our agentic approach leverages verified code snippets from the KB to ensure compilability, while continuously adapting to defensive feedback for robust evasion.

\begin{table}[htbp]
\centering
\caption{EVASION RATE COMPARISON AGAINST SOTA METHODS}
\label{tab:sota_comparison}
\resizebox{\columnwidth}{!}{
\begin{tabular}{l c c c c c c c}
\toprule
\textbf{AV/EDR\textsuperscript{*}} & \textbf{W} & \textbf{T} & \textbf{K} & \textbf{A} & \textbf{M} & \textbf{B} & \textbf{E} \\ \midrule
\textbf{\systemname{}} & \textbf{90.0\%} & \textbf{86.7\%} & \textbf{53.3\%} & \textbf{23.3\%} & \textbf{83.3\%} & \textbf{13.3\%} & \textbf{16.7\%} \\
\boazname{} & 6.9\% & 11.3\% & 0.6\% & 0.0\% & 4.4\% & 0.0\% & 0.0\% \\
\inceptorname{} & 10.0\% & 30.0\% & 0.0\% & 0.0\% & 16.7\% & 0.0\% & 0.0\% \\ 
\dantename{} & 7.1\% & 42.9\% & 7.1\% & 7.1\% & 42.9\% & 0.0\% & 0.0\% \\ 
\bottomrule
\end{tabular}
}
\vspace{2mm}
\parbox{\columnwidth}{\scriptsize \textsuperscript{*}W: Windows Defender, T: Trend Micro, K: Kaspersky, A: Avira, M: McAfee, B: Bitdefender, E: Elastic. \boazname{} (272 generation attempts yielding 159 samples), \inceptorname{} (30 attempts yielding 30 samples) and \dantename{} (300 attempts yielding 14 samples). Evasion rates are calculated based on \textbf{successfully generated samples}, not total generation attempts.}
\end{table}

\begin{figure}[htbp]
\centering
\begin{tikzpicture}
\begin{axis}[
    ybar,
    bar width=5pt, 
    symbolic x coords={0, 1-5, 6-15, 16-25, 26-35, 36-50},
    xtick=data, 
    nodes near coords,
    nodes near coords align={vertical},
    nodes near coords style={font=\tiny, /pgf/number format/.cd, fixed, precision=1, /tikz/.cd, rotate=90, anchor=west}, 
    ymin=0, ymax=65, 
    ylabel={Percentage of Samples (\%)},
    xlabel={VirusTotal Detection Count},
    width=\linewidth,
    height=6cm,
    axis lines*=left,
    ymajorgrids=true,
    grid style=dashed,
    tick label style={font=\footnotesize},
    label style={font=\small},
    enlarge x limits=0.15, 
    legend style={
        at={(0.5, 1.05)}, 
        anchor=south,
        legend columns=-1, 
        font=\footnotesize,
        draw=black, 
        /tikz/every even column/.append style={column sep=0.3cm}
    }
]

\addplot[fill=barred, draw=black] coordinates {
    (0,7.8) (1-5,55.1) (6-15,14.1) (16-25,17.1) (26-35,2.2) (36-50,3.7)
};

\addplot[preaction={fill=baryellow}, pattern=north east lines, pattern color=black, draw=black] coordinates {
    (0,0) (1-5,0) (6-15,5.7) (16-25,36.5) (26-35,57.9) (36-50,0)
};

\addplot[preaction={fill=bargreen}, pattern=grid, pattern color=black, draw=black] coordinates {
    (0,0) (1-5,0) (6-15,26.7) (16-25,20.0) (26-35,53.3) (36-50,0)
};

\addplot[preaction={fill=barblue}, pattern=dots, pattern color=black, draw=black] coordinates {
    (0,0) (1-5,0) (6-15,42.9) (16-25,50.0) (26-35,7.1) (36-50,0)
};

\legend{\systemname{}, \boazname{}, \inceptorname{}, \dantename{}}

\end{axis}
\end{tikzpicture}
\caption{Comparison of VirusTotal Detection Results}
\label{fig:vt_distribution}
\end{figure}
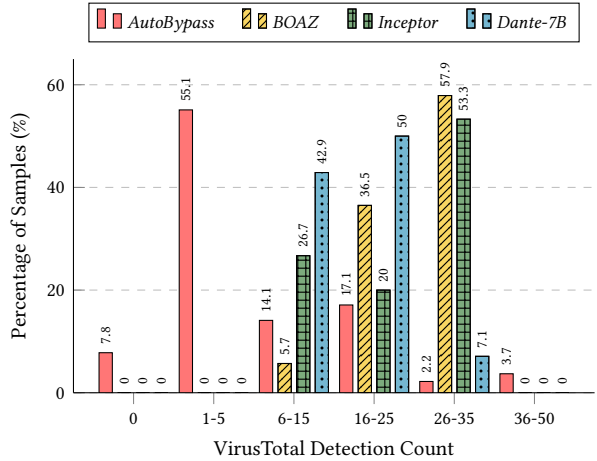

To conduct a broader assessment, we uploaded all (3*7*30=) 630 samples generated by \systemname{} to VirusTotal and benchmarked them against 159 payloads from \boazname{}, 30 from \inceptorname{}, and 14 from \dantename{}~\cite{aghakhani2020malware}. As illustrated in \autoref{fig:vt_distribution}, \systemname{} demonstrates superior stealth, with 62.9\% of samples triggering five or fewer detections. We adopt $\leq$5 detections as the evasion threshold following VirusTotal's own threat hunting guidances~\cite{virustotal_threathunting_2022}. In contrast, all baselines failed to produce any samples within this low-detection range. Instead, over 50\% of \boazname{}, \inceptorname{}, and \dantename{} payloads triggered more than 16 engines, with the majority concentrated in the 16--35 detection range. This result confirms that \systemname{}'s knowledge-driven polymorphism significantly outperforms both static templates and prior LLM-based models in generating evasive payloads.


The architectural foundation enabling this stealth is the KB. Two security experts independently labeled the top-3 results for all 691 unique queries issued during 5,411 total invocations (8 person-hours total), judging whether the returned entries were contextually appropriate for the given query intent. Notably, 82.1\% of queries were rated as retrieving at least one relevant result. The remaining misses were overwhelmingly broad exploration queries used by the \textit{Strategist} for initial surveying rather than precise lookup. For instance, querying ``EDR bypass techniques for shellcode execution'' surfaces \texttt{classic\_loader} as the top result. Even for broad exploration queries without a single exact match, experts consistently rated the returned entries as actionable starting points for semantic orchestration.


\begin{table}[htbp]
\centering
\caption{Evasion Rates Upon 1 Month Retest Under Telemetry}
\label{tab:temporal}
\small
\begin{tabular}{lcccc}
\toprule
\textbf{AV/EDR} & \textbf{0h (Initial)} & \textbf{1 month} \\ \midrule
Windows Defender    & 87.8\%  & 76.7\% \\
Trend Micro AV      & 85.6\%  & 73.3\% \\
Avira      & 43.3\%  & 38.9\% \\
McAfee      & 84.4\%  & 83.3\% \\
Kaspersky EDR       & 55.6\%  & 41.1\% \\
Bitdefender EDR     & 50.0\%  & 37.8\% \\
Elastic Security    & 20.0\%  & 18.9\% \\ \bottomrule
\end{tabular}
\parbox{\linewidth}{\scriptsize
    \textsuperscript{*}Evasion rates are evaluated across (3*30=) 90 runs per AV/EDR ($N = 90$ per cell).
}
\end{table}

Beyond accurate knowledge retrieval, operational stealth also dictates evasion success. Regarding victim process selection, the framework defaults to self-injection, executing the payload within the loader's own address space to avoid cross-process syscalls that behavioral engines routinely flag. For techniques requiring a fresh host process (e.g., \texttt{early\_bird}), the \textit{Strategist} dynamically spawns \texttt{notepad.exe} as a sacrificial process, selected for its ubiquity and low baseline of anomalous child process patterns.

While these operational choices yield high initial evasion rates, modern defenses also leverage telemetry analysis. To evaluate delayed detection driven by cloud telemetry, we maintained successful samples on connected hosts and retested them after one month. As shown in ~\autoref{tab:temporal}, evasion rates declined over time. We attribute this degradation to two primary factors. First, public sandbox submission accelerates signature generation. Manual testing confirmed that samples uploaded to VirusTotal triggered new detections much faster than those that were never uploaded. Second, endpoint agents automatically transmit suspicious files and behavioral traces to vendor clouds for automated analysis, where these cloud engines can execute and profile the uploaded artifacts to generate new rules.

\begin{findingbox}
\textbf{Answer to RQ1:} \systemname{} successfully evades all seven targets, achieving success against Windows Defender (90\%) and Trend Micro (86.7\%), while maintaining superiority over baselines even against aggressive engines like Bitdefender and Elastic. This effectiveness is driven by highly accurate KB retrieval (82.1\% top-3 hit rate) and stealthy operational tactics (e.g., leveraging trusted execution contexts), enabling 62.9\% of samples to trigger $\leq$5 VirusTotal alerts. Although cloud telemetry and sandbox submissions degrade evasion rates over a one-month period, the majority of samples remain undetected, proving the temporal robustness of autonomous semantic adaptation.
\end{findingbox}

\subsection{RQ2: Ablation Study}
\label{subsec:rq2_ablation}

To isolate the impact of our architecture from payload variations, we conducted this ablation study utilizing exclusively the standard Cobalt Strike shellcode. We evaluated our framework under five different configurations using two computationally efficient models (\texttt{Qwen3-8B} and \texttt{Llama-3.1-8B}) against Windows Defender (industry standard baseline) and Kaspersky EDR (more aggressive behavioral inspection). This setup demonstrates that our knowledge-driven approach empowers small baselines independent of the advanced reasoning inherent in large proprietary models. The configurations were: (1) Full KB, (2) No History, (3) No Evasion Techs, (4) No Defensive Rules, and (5) No KB (naive LLM baseline without external knowledge).

\begin{table}[htbp]
\centering
\caption{Ablation Study Results}
\label{tab:ablation_study}
\resizebox{\columnwidth}{!}{
\begin{tabular}{l l l c c c}
\toprule
\textbf{Config} & \textbf{AV/EDR\textsuperscript{*}} & \textbf{Model} & \textbf{C-Succ\textsuperscript{*}} & \textbf{Func\textsuperscript{*}} & \textbf{Succ. Evasion} \\ \midrule
\multirow{4}{*}{Full KB} & \multirow{2}{*}{W} & Qwen3-8B & 29 & 28 & 83.3\% (25/30) \\
 & & Llama-3.1-8B & 29 & 27 & 80.0\% (24/30) \\ \cmidrule{2-6}
 & \multirow{2}{*}{K} & Qwen3-8B & 26 & 26 & 43.3\% (13/30) \\
 & & Llama-3.1-8B & 29 & 28 & 43.3\% (13/30) \\ \midrule
\multirow{4}{*}{\makecell[l]{No History}} & \multirow{2}{*}{W} & Qwen3-8B & 27 & 25 & 83.3\% (25/30) \\
 & & Llama-3.1-8B & 29 & 23 & 76.7\% (23/30) \\ \cmidrule{2-6}
 & \multirow{2}{*}{K} & Qwen3-8B & 27 & 26 & 43.3\% (13/30) \\
 & & Llama-3.1-8B & 29 & 26 & 33.3\% (10/30) \\ \midrule
\multirow{4}{*}{\makecell[l]{No \\ Evasion Techs}} & \multirow{2}{*}{W} & Qwen3-8B & 26 & 22 & 66.7\% (20/30) \\
 & & Llama-3.1-8B & 21 & 10 & 26.7\% (8/30) \\ \cmidrule{2-6}
 & \multirow{2}{*}{K} & Qwen3-8B & 27 & 22 & 40.0\% (12/30) \\
 & & Llama-3.1-8B & 14 & 9 & 30.0\% (9/30) \\ \midrule
\multirow{4}{*}{\makecell[l]{No \\ Defensive Rules}} & \multirow{2}{*}{W} & Qwen3-8B & 19 & 18 & 60.0\% (18/30) \\
 & & Llama-3.1-8B & 24 & 11 & 36.7\% (11/30) \\ \cmidrule{2-6}
 & \multirow{2}{*}{K} & Qwen3-8B & 26 & 24 & 43.3\% (13/30) \\
 & & Llama-3.1-8B & 19 & 18 & 40.0\% (12/30) \\ \midrule
\multirow{4}{*}{No KB} & \multirow{2}{*}{W} & Qwen3-8B & 27 & 16 & 53.3\% (16/30) \\
 & & Llama-3.1-8B & 22 & 10 & 33.3\% (10/30) \\ \cmidrule{2-6}
 & \multirow{2}{*}{K} & Qwen3-8B & 26 & 18 & 36.7\% (11/30) \\
 & & Llama-3.1-8B & 23 & 16 & 26.7\% (8/30) \\ \bottomrule
\end{tabular}
}
\vspace{1mm}
\parbox{\linewidth}{\scriptsize \textsuperscript{*}C-Succ: Compilation Success. Func: Functionality Check Passed. W: Windows Defender. K: Kaspersky EDR. Each experiment consists of 30 iterations.}
\end{table}

The results, summarized in~\autoref{tab:ablation_study}, confirm that the KB is the primary driver of evasion success. Removing the KB entirely drops \texttt{Llama-3.1-8B}'s success against Windows Defender from 80\% to 33.3\%, revealing that LLMs alone lack the domain-specific logic to navigate complex defense engines. Drilling down into individual components, \textit{Evasion \& Encode Techs} are crucial for functional correctness. Removing this module critically impairs the agents' ability to produce executable code. Notably, for \texttt{Llama-3.1-8B} against Windows Defender, the number of functionally valid samples dropped from 27 to 10, causing the final success rate to fall to just 26.7\%. This highlights the KB's role in grounding LLM synthesis and preventing logical hallucinations.

Even when code compiles, \textit{Defensive Rules} and OPSEC constraints are essential for operational stealth. Without this guidance, agents default to conspicuous implementations, such as allocating RWX memory, which increases the forensic footprint and triggers detection. Consequently, \texttt{Qwen3-8B}'s success rate against Windows Defender fell from 83.3\% to 60\% when rules were disabled. Finally, \textit{Historical Experience} drives iterative efficiency. Disabling this component removes the \textit{Strategist}'s ability to learn from past failures, leading to repetitive, easily detectable behaviors. This is reflected in a drop in success rates, with \texttt{Llama-3.1-8B}'s performance against Windows Defender decreasing from 80\% to 76.7\%.

\begin{findingbox}
\textbf{Answer to RQ2:} The KB is the decisive factor for evasion success. Without it, both 8B parameter models degrade significantly. Specifically, \texttt{Llama-3.1-8B} drops from 80\% to 33.3\% against Windows Defender, and \texttt{Qwen3-8B} from 43.3\% to 36.7\% against Kaspersky. Among KB components, \textit{Evasion Techs} is paramount for functional correctness, as removing it collapses Llama's valid samples from 27 to 10, while \textit{Defensive Rules} ensures operational stealth, evidenced by \texttt{Qwen3-8B} dropping from 83.3\% to 60.0\% without it. This confirms the KB functions as an intelligence amplifier that elevates small models to practical evasion capability.
\end{findingbox}

\subsection{RQ3: Model Sensitivity}
\label{subsec:rq3_model_sensitivity}

Maintaining experimental consistency, we deployed only the Cobalt Strike payload to evaluate framework sensitivity to the underlying LLM. We tested seven models, ranging from large proprietary models (\texttt{GPT-5.1}) to smaller open-weight models (\texttt{Qwen3-8B}, \texttt{Llama-3.1-8B}). \autoref{tab:rq3_model_comparison} summarizes the results.


\begin{table}[htbp]
\centering
\caption{Model Sensitivity Analysis}
\label{tab:rq3_model_comparison}
\scriptsize
\renewcommand{\arraystretch}{0.85}
\resizebox{\columnwidth}{!}{
\begin{tabular}{l l c c c}
\toprule
\textbf{Model} & \textbf{AV/EDR\textsuperscript{*}} & \textbf{Succ. Evasion} & \textbf{Avg. Token (In/Out)\textsuperscript{*}} & \textbf{Price (\$)\textsuperscript{*}}\\ \midrule
\multirow{5}{*}{GPT-5.1} & W & 90.0\% (27/30) & 7.2k / 2.3k & 0.032\\
 & T & 86.7\% (26/30) & 9.5k / 3.1k & 0.043 \\
 & K & 53.3\% (16/30) & 15.2k / 4.5k & 0.064 \\
 & B & 13.3\% (4/30) & 70.1k / 18.1k & 0.269 \\
 & E & 16.7\% (5/30) & 51.9k / 12.7k & 0.192 \\
 \midrule
\multirow{5}{*}{DeepSeek-R1} & W & 83.3\% (25/30) & 13.1k / 10.2k & 0.032 \\
 & T & 83.3\% (25/30) & 14.2k / 9.4k & 0.032 \\
 & K & 56.7\% (17/30) & 20.9k / 14.3k & 0.045 \\
 & B & 10.0\% (3/30) & 139.1k / 75.3k & 0.271 \\
 & E & 23.3\% (7/30) & 53.4k / 35.3k & 0.120 \\
 \midrule
\multirow{5}{*}{DeepSeek-V3} & W & 80.0\% (24/30) & 12.2k / 1.4k & 0.0058 \\
 & T & 90.0\% (27/30) & 11.0k / 1.2k & 0.0043\\
 & K & 53.3\% (16/30) & 17.7k / 2.2k & 0.0072\\
 & B & 0.0\% (0/30) & N/A & N/A \\
 & E & 0.0\% (0/30) & N/A & N/A \\
 \midrule
\multirow{5}{*}{Gemini-2.5-Flash} & W & 73.3\% (22/30) & 14.0k / 1.6k & 0.008 \\
 & T & 66.7\% (20/30) & 14.9k / 1.7k & 0.009\\
 & K & 56.7\% (17/30) & 17.3k / 2.0k & 0.010\\
 & B & 3.3\% (1/30) & 285.7k / 34.1k & 0.171 \\
 & E & 6.7\% (2/30) & 158.4k / 17.8k & 0.092 \\
 \midrule
\multirow{5}{*}{Llama-3.1-70B} & W & 66.7\% (20/30) & 11.4k / 2.2k & 0.002 \\
 & T & 90.0\% (27/30) & 9.6k / 1.9k & 0.002 \\
 & K & 46.7\% (14/30) & 17.4k / 3.4k & 0.003 \\
 & B & 0.0\% (0/30) & N/A & N/A \\
 & E & 6.7\% (2/30) & 114.9k / 28.0k & 0.057 \\
 \midrule
\multirow{5}{*}{Llama-3.1-8B} & W & 80.0\% (24/30) & 12.3k / 2.3k & 0.0005 \\
 & T & 80.0\% (24/30) & 14.1k / 2.0k & 0.0005 \\
 & K & 43.3\% (13/30) & 26.7k / 4.4k & 0.0010 \\
 & B & 0.0\% (0/30) & N/A & N/A\\
 & E & 0.0\% (0/30) & N/A & N/A\\
 \midrule
\multirow{5}{*}{Qwen3-8B} & W & 83.3\% (25/30) & 17.8k / 1.2k & 0.0016 \\
 & T & 80.0\% (24/30) & 20.0k / 1.4k & 0.0018 \\
 & K & 43.3\% (13/30) & 36.0k / 2.4k & 0.0032 \\
 & B & 0.0\% (0/30) & N/A & N/A \\
 & E & 0.0\% (0/30) & N/A & N/A \\
 \midrule
\end{tabular}
}
\parbox{\linewidth}{\scriptsize \textsuperscript{*}W: Windows Defender, T: Trend Micro, K: Kaspersky, B: Bitdefender, E: Elastic. Avg. Token represents the average token cost per successful evasion. Prices from \url{https://pricepertoken.com/} (Feb 2, 2026)}
\end{table}

Our evaluation yields two key insights. First, although larger models like \texttt{GPT-5.1} show superior performance and cost efficiency, smaller open-weight models remain effective. For example, \texttt{DeepSeek-V3} and \texttt{DeepSeek-R1} achieved over 50\% success against multiple targets, proving our framework functions without relying exclusively on state-of-the-art models.

Second, the performance gap between large and small models is narrower than in general code generation. Small models like \texttt{Llama-3.1-8B} still achieve successful evasion because our KB offloads complex security reasoning. By providing verified snippets, OPSEC constraints, and historical feedback, the KB simplifies the LLM's task from innovating novel evasion logic to orchestrating preexisting high-quality techniques.

However, tradeoffs exist in size, cost, and capability. Smaller models successfully formulate strategies but struggle with tactical implementation, causing more build failures that necessitate \textit{Debugger Agent} intervention. Conversely, larger models like \texttt{GPT-5.1} consistently produce compilable C++ code on the first attempt.

\begin{findingbox}
\textbf{Answer to RQ3:} The framework exhibits strong robustness across LLM scales. Against Windows Defender, 8B models (\texttt{Qwen3-8B}: 83.3\%, \texttt{Llama-3.1-8B}: 80\%) approach \texttt{GPT-5.1}'s 90\% at 20 to 60 times lower cost. Unlike typical code generation, our KB simplifies the task to technique orchestration, allowing smaller models to rival SOTA performance. The tradeoff is increased compilation failures requiring more \textit{Debugger} iterations.
\end{findingbox}

\subsection{RQ4: Analysis of Evasion Techniques}
\label{subsec:rq4_techniques}

A key advantage of \systemname{} is its ability to systematically evaluate the effectiveness of individual evasion techniques at scale. To achieve this, we aggregate the execution logs from all seven models tested in \autoref{subsec:rq3_model_sensitivity}. The results, summarized in~\autoref{tab:technique_effectiveness}, provide a quantitative measure of each technique's success rate, accounting for compilation failures, execution failures, and EDR alerts.

\begin{table}[htbp]
\centering
\caption{Effectiveness of Individual Evasion Techniques}
\label{tab:technique_effectiveness}
\footnotesize 
\renewcommand{\arraystretch}{1.0} 
\resizebox{\columnwidth}{!}{
\begin{tabular}{l c c c c r}
\toprule
\textbf{Technique} & \textbf{C-Fail\textsuperscript{*}} & \textbf{E-Fail\textsuperscript{*}} & \textbf{Alert} & \textbf{Succ. Evasion} \\ \midrule
\texttt{module\_overload} & 0 & 0 & 9 &  74.3\% (26/35) \\
\texttt{remote\_thread\_hijack} & 0 & 4 & 18 &  62.1\% (36/58) \\
\texttt{fls\_callback} & 0 & 0 & 14 & 60.0\% (21/35) \\
\texttt{section\_injection} & 5 & 16 & 84 & 57.5\% (142/247) \\
\texttt{threadless} & 2 & 15 & 56 & 56.5\% (95/168) \\
\texttt{classic} & 0 & 12 & 44 & 55.6\% (70/126) \\
\texttt{ekko\_sleep} & 1 & 9 & 67 & 55.2\% (95/172) \\
\texttt{etw\_patch} & 9 & 27 & 173 & 54.2\% (247/456) \\
\texttt{callback\_trigger} & 0 & 12 & 49 & 52.0\% (66/127) \\
\texttt{unhook\_ntdll} & 9 & 27 & 139 & 51.4\% (185/360) \\
\texttt{amsi\_bypass} & 4 & 20 & 137 & 51.2\% (169/330) \\
\texttt{dirty\_vanity} & 1 & 10 & 32 & 50.6\% (44/87) \\
\texttt{queue\_user\_apc} & 0 & 10 & 33 & 50.0\% (43/86) \\
\texttt{hells\_gate} & 3 & 17 & 90 & 49.1\% (106/216) \\
\texttt{register\_wait} & 3 & 14 & 55 & 48.9\% (69/141) \\
\texttt{fiber\_injection} & 8 & 24 & 108 & 48.0\% (129/269) \\
\texttt{thread\_pool} & 3 & 24 & 29 & 47.7\% (51/107) \\
\texttt{uuid\_conversion} & 0 & 5 & 31 & 45.5\% (30/66) \\
\texttt{early\_bird} & 0 & 7 & 33 & 41.2\% (28/68) \\
\bottomrule
\end{tabular}
}
\parbox{\linewidth}{\scriptsize \textsuperscript{*}C-Fail: Compilation Failure, E-Fail: Execute Failure.}
\end{table}

Our analysis reveals that evasion success is heavily influenced by the technique's ability to mimic legitimate system behavior. \texttt{module\_overload} (74.3\%) and \texttt{remote\_thread\_hijack} (62.1\%) demonstrated the highest effectiveness by masking malicious memory as disk-backed sections or abusing less-monitored Fiber Local Storage mechanisms. In contrast, telemetry-blinding techniques like \texttt{etw\_patch} and \texttt{unhook\_ntdll} incurred the highest alert volumes (173 and 139), indicating that modern EDRs actively monitor the integrity of their own hooks. The results also highlight a trade-off between complexity and reliability: techniques involving runtime memory modification, such as \texttt{fiber\_injection}, suffered high execution failure rates (24), while legacy patterns like \texttt{early\_bird} (41.2\%) have been largely fingerprinted by modern behavioral engines.


\begin{table}[htbp]
\centering
\caption{Evasion Success Rate by Sample Type}
\label{tab:format_effectiveness}
\resizebox{\columnwidth}{!}{
\begin{tabular}{l c c c c c c}
\toprule
\textbf{AV/EDR\textsuperscript{*}} & \textbf{W} & \textbf{T} & \textbf{K} & \textbf{B} & \textbf{E} & \textbf{Overall} \\ \midrule
Standalone EXE 
& \makecell{90/117\\76.9\%} 
& \makecell{90/109\\82.6\%} 
& \makecell{11/98\\11.2\%} 
& \makecell{2/104\\1.9\%} 
& \makecell{5/102\\4.9\%} 
& \makecell{198/530\\37.4\%} \\
\addlinespace
DLL Sideloading 
& \makecell{77/93\\82.8\%} 
& \makecell{83/101\\82.2\%} 
& \makecell{95/112\\84.8\%} 
& \makecell{6/106\\5.7\%} 
& \makecell{11/108\\10.2\%} 
& \makecell{272/520\\52.3\%} \\
\bottomrule
\end{tabular}
}
\vspace{1mm}
\parbox{\columnwidth}{\scriptsize \textsuperscript{*}W: Windows Defender, T: Trend Micro, K: Kaspersky EDR, B: Bitdefender EDR, E: Elastic Security.}
\end{table}

As summarized in~\autoref{tab:format_effectiveness}, which aggregates all samples generated by the seven models in~\autoref{tab:rq3_model_comparison}, we also observed a clear performance disparity based on the execution context. Techniques that operate within a separate, trusted process, such as those involving DLL Sideloading, consistently demonstrated higher success rates than standalone executables. This is because loading malicious code into a legitimate process's address space effectively inherits a degree of trust, making the activity appear less anomalous to behavioral heuristics.

\begin{findingbox}
\textbf{Answer to RQ4:} Our analysis of 19 techniques reveals that evasion success depends more on execution context than technical sophistication. DLL Sideloading achieves 52.3\% success versus 37.4\% for standalone EXEs, as inheriting a legitimate process's trust evades behavioral heuristics. Among individual techniques, mimicry-based approaches like \texttt{module\_overload} achieve up to 74.3\% success, outperforming aggressive tampering like \texttt{etw\_patch}, which triggers heavy alert volumes (173) due to EDR self-integrity monitoring. Complex techniques (e.g., \texttt{fiber\_injection}) suffer execution instability, while legacy patterns like \texttt{early\_bird} drop to 41.2\% as modern EDRs now recognize these signatures.
\end{findingbox}




\section{Related Work}
\label{sec:related_work}

\subsection{EDR Evasion Techniques}

EDR evasion research spans hook bypass, memory stealth, telemetry blinding, and execution context abuse. At the API interception layer, \textit{HookChain}~\cite{junior2024hookchain} redirects execution flow to circumvent user-mode hooking. Similarly, indirect syscall techniques directly invoke kernel services to avoid \texttt{ntdll.dll} instrumentation entirely~\cite{hand2023evading}. To counter memory scanning, sleep obfuscation techniques such as Ekko and Cronos encrypt the in-memory beacon image during idle intervals and restore it prior to execution. This approach targets the periodic scan window exploited by EDR memory scanners~\cite{dobin_edr}. Stack spoofing complements these techniques by forging call stack frames to defeat call chain based heuristics. At the telemetry layer, ETW patching and AMSI bypass techniques blind the sensor collection pipeline before payload execution. Beyond technique-level work, \textit{EvilEDR}~\cite{alachkar2025eviledr} demonstrates that EDR agents themselves can be repurposed as offensive tools to fundamentally invert the defender-attacker asymmetry. At a higher level, mimicry attacks~\cite{goyal2023sometimes} embed malicious actions within benign provenance subgraphs. ANIMAGUS~\cite{zhou2023limits} also mimics legitimate I/O patterns to evade ransomware-specific detectors. While individually effective, these techniques address isolated detection vectors and require significant manual expertise to compose across diverse EDR products.

\subsection{Automated Malware Generation}

Existing automation approaches span binary mutation, template-based generation, and LLM-assisted synthesis. At the binary level, AIMED~\cite{castro2019aimed} applies genetic programming to mutate malware binaries. MAB-Malware~\cite{song2022mab} formulates evasion as a multi-armed bandit problem. However, Pierazzi et al.~\cite{pierazzi2020intriguing} demonstrate that problem-space adversarial examples must preserve file validity and execution semantics. Feature-space perturbations routinely ignore this critical constraint. At the source level, template-based frameworks like \boazname{}~\cite{boaz} and \inceptorname{}~\cite{inceptor} automate shellcode loader generation by inserting payloads into predefined code skeletons. Their reliance on static structure produces recognizable fingerprints that EDRs quickly learn to detect. Most recently, \dantename{}~\cite{dante7b} fine-tunes a 7B model on malware development data. This approach achieves only a 4.7\% compilation success rate, underscoring that raw LLM capability without structured domain knowledge is insufficient for reliable exploit synthesis. Critically, all existing approaches operate in an open-loop manner and fail to incorporate live EDR feedback to drive iterative strategy refinement. \systemname{} addresses this gap through semantic-level KB guided reasoning and closed-loop alert inference to enable both polymorphic generation and adaptive strategy evolution.

\subsection{LLMs for Offensive Security}
LLMs have been applied to security tasks including fuzzing and vulnerability discovery. TitanFuzz~\cite{deng2023large} and Fuzz4All~\cite{xia2024fuzz4all} leverage LLMs to generate input programs for fuzzing, while PentestGPT~\cite{deng2024pentestgpt} orchestrates LLMs to guide penetration testing workflows. However, these focus on \textit{identifying vulnerabilities} rather than \textit{weaponizing artifacts} against active defenses. Furthermore, Sandoval et al.~\cite{sandoval2023lost} showed unguided LLM code often contains functional defects or triggers immediate detection. Crucially, existing LLM approaches operate open-loop, generating samples without validating against live defenses. \systemname{} closes this gap through a closed-loop architecture where the \textit{Tester} infers detection root causes from system telemetry, feeding insights back to drive evolutionary strategy refinement.

\section{Discussion}
\label{sec:discussion}

\noindent\textbf{Limitations.}
\systemname{} relies on LLM variability to generate diverse code structures, effectively evading defenses through high-level strategic iteration. However, it does not yet replicate the fine-grained tradecraft employed by human experts. When facing static detection, analysts often use binary splitting (e.g., VirTest) to isolate the exact byte sequence triggering a signature. For behavioral alerts, experts deploy debugging utilities such as \textit{x64dbg}~\cite{x64dbg} to pinpoint the specific API call causing the block. Currently, \systemname{} addresses detection failures by regenerating entire modules rather than performing surgical modifications. Future work will introduce a dedicated localization stage to automate these precise refinement tasks. 


We acknowledge our alert reasoning module relies on heuristic inference. Because commercial endpoint protections emit opaque alerts, our framework deduces root causes empirically by monitoring objective OS artifacts. While highly effective, this cannot guarantee absolute attribution accuracy.

Finally, restricting our deployment scope exclusively to shellcode loaders constitutes a notable limitation. Although this focus provides a robust baseline, it omits other prevalent attack vectors, such as script execution environments (PowerShell, WMI). Expanding our orchestration to encompass these diverse vectors remains a critical direction for future research.

\noindent\textbf{Generality and Adaptation.}
A key design goal of \systemname{} is broad generality across diverse endpoint protections with varying alert formats and detection mechanisms. This adaptability stems from two core principles. First, our pipeline possesses autonomous learning capabilities, continuously accumulating successful evasion patterns for each specific target through iterative historical feedback. Second, the evaluation module remains entirely agnostic to proprietary interfaces. It infers evasion outcomes purely by monitoring universal operating system phenomena, such as process survival and network traffic establishment. By relying on objective system states rather than proprietary telemetry, the framework scales across endpoint defenses without manual customization.

\noindent\textbf{Implications for Defense.}
Our evaluation reveals several insights for AV/EDR vendors. First, the high success rates against Windows Defender (90\%) and Trend Micro (86.7\%) suggest that signature-based and basic behavioral detection remain insufficient against polymorphic, knowledge-driven attacks. Second, as demonstrated in RQ4, techniques leveraging trusted execution contexts (e.g., DLL sideloading) consistently achieve higher evasion rates across all tested EDRs. This indicates a systemic blind spot: current defenses struggle to distinguish malicious code executing within legitimate process contexts from benign operations. We recommend that vendors enhance monitoring of code injection into signed binaries and implement stricter validation of DLL loading sequences, even for trusted applications.

\noindent\textbf{Responsible Disclosure.}
We have disclosed our findings and shared representative samples with all seven AV/EDR vendors evaluated in this study. At the time of submission, we have received acknowledgment from two vendors, confirming that our reported samples have been incorporated into their detection pipelines. We will coordinate with the remaining vendors before any public release of technical details.
\section{Conclusion}
\label{sec:conclusion}

We presented \systemname{}, to our knowledge the first framework to model EDR evasion as a knowledge-driven, feedback-directed agentic process. By integrating a Detection-Aware Knowledge Base with multi-agent orchestration, our approach achieves 90\% and 86.7\% evasion rates against Windows Defender and Trend Micro respectively, and our ablation study confirms that the KB elevates small open-weight models to match large proprietary models. Our technique analysis further reveals that trusted execution contexts, such as DLL sideloading, consistently evade modern defenses by inheriting the benign reputation of legitimate processes, exposing a systemic blind spot. Beyond immediate findings, evaluating seven commercial products demonstrates that the knowledge required to evade modern defenses is already public and systematically operationalizable. We hope \systemname{} provides a foundation for continuous endpoint protection evaluation, encouraging vendors to proactively harden identified attack surfaces.

\section*{Ethical Considerations}

\textbf{Stakeholder Analysis.} The publication of this framework impacts multiple distinct groups. EDR vendors and enterprise security teams stand to benefit significantly by gaining a realistic benchmark to identify systemic blind spots before they are exploited in the wild. Conversely, enterprise networks and everyday end users face potential negative impacts if threat actors are motivated by our findings to utilize artificial intelligence for malware evolution. We acknowledge the concern that demonstrating these capabilities might accelerate the adversarial adoption of automated evasion techniques.

\textbf{Mitigations and Advance Notification.} To ensure the defensive benefits outweigh the potential risks, we implemented strict mitigations. Most crucially, we notified the vendors of all seven analyzed endpoint protection systems well in advance of submission. We provided them with complete reports, inferred root causes, and actionable defensive recommendations. This advance disclosure ensures that defenders have the necessary time to upgrade their telemetry and detection pipelines against AI evolved threats before such automated methodologies become broadly mainstream.

\textbf{Limited Release Strategy.} To further mitigate the risk of direct weaponization while upholding the principles of open science, we have adopted a limited release strategy. We are making the core multi agent orchestration logic publicly available. However, we will strictly withhold the complete pre-populated offensive knowledge base. Instead, to ensure the framework remains fully functional and verifiable for legitimate academic research, we will provide a carefully curated minimal subset containing three of the least effective evasion techniques. This calculated balance allows researchers to validate our methodology and reproduce the workflow without providing unskilled attackers with a ready to use offensive arsenal.

\textbf{Justification for Publication.} Ultimately, our decision to publish is driven by the conviction that the current gap in automated EDR evaluation poses a severe risk. While manual evasion techniques are already prevalent in the wild, the security community lacks a scalable method to stress test defenses against them. By introducing this framework and sharing our findings responsibly, we provide a rigorous benchmark for EDR resilience. We believe this controlled disclosure is in the best interest of the public, as it catalyzes the essential hardening of the endpoint security ecosystem against future automated threats.

\section*{Artifacts}

To support reproducible research, we have made the core architecture of \systemname{} publicly available. However, due to the sensitive nature of automated bypass and the potential for malicious misuse, we have adopted a limited release strategy. We explicitly acknowledge that withholding the complete pre-populated knowledge base reduces the direct reproducibility of our core claims. Instead, we release the complete multi-agent orchestration logic, testing infrastructure, and a minimal subset containing three of the least effective evasion techniques for academic verification. These artifacts are available in our anonymized repository: \url{https://anonymous.4open.science/r/bypass-agent-9BE6}.

\bibliographystyle{ACM-Reference-Format}
\bibliography{ref}

\end{document}